\documentstyle[12pt]{article}
\newbox\rotbox
\def\be{\begin{eqnarray}}
\def\ee{\end{eqnarray}}
\def\MeV{\nobreak\,\mbox{MeV}}
\def\GeV{\nobreak\,\mbox{GeV}}

\def\qbar{\overline{q}}
\def\ubar{\overline{u}}
\def\dbar{\overline{d}}
\def\sbar{\overline{s}}

\def\bra#1{\langle #1|}
\def\ket#1{| #1\rangle}
\def\nucbra{\bra{ p}}
\def\nucbrap{\bra{ p'}}
\def\nucket{\ket{ p}}

\begin{document}

\vglue 2 cm
\centerline{\bf STRANGENESS IN THE NUCLEON: } 
\vskip 3 mm
\centerline{\bf THE STRANGE VECTOR FORM FACTORS }

\vskip 1 cm

\centerline{Hilmar Forkel}
\vskip 0.5cm

\centerline{European Centre for Theoretical Studies in Nuclear 
Physics and Related Areas, }
\centerline{Villa Tambosi, Strada delle Tabarelle 286, I-38050 
Villazzano, Italy}

\vskip 1cm
\vskip  2\baselineskip
\hskip 2.5cm ABSTRACT

We discuss two descriptions of the nucleon's strange vector form 
factors in the framework of vector meson dominance. The first, an 
updated and extended version of Jaffe's dispersion analysis, 
approximates the spectral functions of the form factors as a sum 
of vector meson poles, whereas the second combines vector meson 
dominance in the $\omega$ and $\phi$ meson sector with an intrinsic 
strangeness distribution from a kaon cloud.

\vskip 2\baselineskip
\hskip 2.5cm KEYWORDS

Nucleon properties; strangeness; strange quark; dispersion theory; 
vector meson dominance.

\vskip 2\baselineskip
\hskip 2.5cm STRANGENESS IN THE NUCLEON
\label{intro}

The strangeness content of the nucleon, as probed by the matrix 
elements $\bra{N} \sbar \Gamma s \ket{N} $ of strange quark operators 
$s, \bar{s}$ in a Lorentz channel specified by $\Gamma$, 
offers a key to intriguing and little understood quantum effects in the 
nucleon wave function. Due to the relatively small strange quark mass 
these effects can be sizable, and growing experimental evidence 
indeed indicates unexpectedly large amounts of strangeness in the 
nucleon. From pion-nucleon scattering data, for example, one can 
extract the ratio 
\be
R_s = {\nucbra \sbar s\nucket\over\nucbra \ubar u + \dbar d
+ \sbar s\nucket} \; \label{r}
\ee
($u$, $d$ and $s$ are the up, down and strange quark fields, and 
$\nucket$ denotes the nucleon state), and obtains surprisingly large 
(although somewhat controversial) values, up to $R_s\simeq 0.1 - \, 
0.2$ (Cheng and Dashen, 1971; Cheng, 1976; Donoghue and Nappi, 1986; 
Gasser, Leutwyler and Sainio, 1991; Kluge, 1995). 
This implies that $\nucbra \sbar s\nucket$ can reach almost half the 
magnitude of the corresponding up-quark matrix element and that the 
nucleon mass would be reduced by $\approx 300 \MeV$ in a world with 
massless strange quarks.

Further and more direct evidence for sizeable strange quark effects 
in the nucleon has emerged since the end of the eighties from deep 
inelastic $\mu$-$p$ scattering data. The European Muon Collaboration 
(EMC) measured the polarized proton structure function $g_1^p(x)$ in 
a large range of the Bjorken variable, $x \in [0.01,0.7]$ (Ashman 
{\it et al.}, 1988, 1989) and found, after Regge extrapolation to 
$x=0$ and combination with earlier SLAC data,  
\be
\int_0^1 dx \, g_1^p(x) = 0.126\pm 0.010 \, (stat) \pm 0.015 \, 
(syst) \; 
\ee
at $Q^2 = 10 \GeV^2 / c^2$. Without strange quark contributions 
one would expect, following Ellis and Jaffe (1974), a significantly 
larger value, $0.175 \pm 0.018$. The data therefore indicate a 
nonvanishing strange quark contribution $\Delta s = -0.16\pm 0.008$ 
to the proton spin (if $SU(3)$ is not too badly broken), or 
equivalently, via the Bjorken sum rule, a substantial 
strangeness contribution $\nucbra \sbar \gamma_\mu \gamma_5 s\nucket$ 
to the proton matrix element of the isoscalar axial-vector current. 
The low-energy elastic $\nu$--$p$ 
scattering experiment E734 at Brookhaven (Ahrens {\it et al.}, 1987) 
complemented the EMC data by measuring the same matrix element at 
smaller momenta $(0.4 \GeV^2 < Q^2 < 1.1 \GeV^2)$. The extracted 
axial vector current form factors are consistent with the muon 
scattering data.

The above experimental findings indicate a role of strange quarks in 
the nucleon that goes beyond naive quark model expectations (for a more 
complete discussion see (Alberg, 1995)) and have triggered further 
theoretical and experimental investigations. In view of the expected 
channel-dependence of the  strange quark matrix elements (see, for 
example, (Ioffe and Karliner, 1990)), an important part of this 
activity is directed towards new channels and, in particular, to the 
vector channel. The vector current matrix element describes the 
nucleon's strangeness charge and current distributions (in analogy to 
the electromagnetic case) by Dirac and Pauli form factors, 
\be
\nucbrap  \sbar \gamma_\mu s \nucket = \overline{N}(p^\prime) 
\left(\gamma_\mu F_{1}^{(s)} (q^2) +
i\frac{\sigma_{\mu\nu}q^\nu} {2M_N} F_{2}^{(s)} (q^2)
\right) N(p). \label{sff}
\ee
($N$ is the free Dirac spinor of the nucleon and $q = p^\prime - p$.) 
Particularly attractive features of the vector matrix element are its 
scale independence (due to strangeness conservation, i.e. up to weak 
corrections) and its direct experimental accessibility via 
parity-violating lepton scattering off different hadronic targets. 
Several experiments of this type are in preparation at CEBAF and MAMI 
(Musolf {\it et al.}, 1994), and SAMPLE at Bates (McKeown and 
Beck, 1989) already started to take data. These experiments will, 
in fact, provide the first direct low-energy measurements of sea quark 
effects in hadrons. 

The expected data will determine, in particular, the leading 
nonvanishing moments of the vector strangeness distribution, 
namely the strangeness radius $r^2_s$ and magnetic moment $\mu_s$, 
\be
\mu_s = F_2^{(s)} (0) = G_M^{(s)}(0), \qquad r^2_{s} = 6 \frac{d }{d q^2} 
F_1^{(s)} (q^2) |_{q^2=0}, \qquad (r^2_{s})_{Sachs} =  6 \frac{d }{d q^2} 
G_E^{(s)} (q^2) |_{q^2=0}. \label{mur}
\ee
Note the two alternative definitions of the moments, which are both
currently in use. One of them is based on the Sachs form factors
\be
G_E^{(s)}(q^2) = F_1^{(s)} (q^2) + {q^2\over4M_N^2}F_2^{(s)} (q^2), 
\qquad  G_M^{(s)} (q^2) = F_1^{(s)} (q^2) + F_2^{(s)} (q^2) .
\label{sachs}
\ee
Since sea quark distributions in hadrons arise from a subtle 
interplay of quantum effects in QCD, their reproduction in hadron 
models is much more challenging than the calculation of the standard 
static observables. Reflecting these difficulties, present model 
calculations of the strange form factors (for a comparison see 
(Forkel {\it et al.}, 1994)) contain large and often uncontrolled 
theoretical uncertainties and are partially inconsistent with each 
other. Lattice calculations of strange sea quark distributions, on 
the other hand, are computationally very demanding and have not yet 
been carried out (see, however, (Liu and Dong, 1994)). The particular 
value of the strange quark mass, which is neither light nor heavy 
compared to the QCD scale $\Lambda_{QCD}$, further complicates the 
theoretical 
situation. In contrast to the light up and down quarks, the effects of 
the heavier strange quark are much harder to approach from the chiral 
limit, {\it i.e.} by an expansion in the quark mass. On the other hand, 
the strange quark is too light for the methods of the heavy-quark 
sector, {\it e.g.} the nonrelativistic approximation or the heavy-quark 
symmetry, to work. 

In the following we discuss two theoretical approaches to the strange 
form factors which bypass, to a different degree, the need for detailed 
nucleon model calculations. They are both based on the phenomenologically 
successful vector meson dominance (VMD) concept, which is implemented in 
the following section in a dispersion theoretical framework (Jaffe, 1989; 
Forkel, 1995), and in the subsequent section via current field 
identities (Forkel {\it et al.}, 1994).

\vskip 2\baselineskip
\hskip 2.5cm DISPERSION ANALYSIS

The dispersive approach, initiated by Jaffe (1989), permits a 
nucleon-model independent estimate of the strange form factors on 
the basis of phenomenological input. It starts from the dispersion 
relations 
\be
F_i^{(s)} (q^2) = \frac{1}{\pi} \int_{s_0}^\infty d s \, 
\frac{Im \{F_i^{(s)}(s)\}}{s-q^2},
\ee
where $s_0 = (3 m_{\pi})^2$ is the three-pion threshold and subtraction
terms are suppressed. The spectral functions $\pi^{-1} Im \{F_i^{(s)}\}$ 
receive contributions from intermediate (on-shell) states with 
$ I^G \, J^{PC} = 0^- 1^{- -}$, through which the strangeness current 
couples to the nucleon. In the pole approximation they are represented 
by $N$ sharp vector meson states,
\be
\frac{1}{\pi} Im \{F_i^{(s)}(s) \} = \sum_{v=1}^N B_i^{(v)} m_v^2 \,
\delta (s - m_v^2), \qquad i \in \{1,2 \}. \label{Npsd}
\ee
This ansatz is perfectly adequate for the two lowest-lying, narrow-width 
resonances $\omega$ and $\phi$. The additional poles are effectively 
summarizing strength from higher lying, broader resonances and from 
continuum contributions. Eq. (\ref{Npsd}) contains $3 N$ a priori 
unknown mass and coupling parameters. Jaffe (1989) realized that the 
three lowest-lying masses and the couplings of the $\omega$ and $\phi$ 
poles can be estimated model-independently, since (i) there exists 
another current, the isoscalar electromagnetic current $J^{(I=0)}_\mu$, 
which carries the same quantum numbers as the strange 
current and thus couples through the same intermediate states, (ii) the 
associated isoscalar form factors are well measured in a large range of 
$q^2$ and well fitted by a dispersive 3-pole ansatz (H{\"o}hler {\it et 
al.}, 1976; Mergell {\it et al.}, 1995) and (iii) the flavor structure 
of the first two poles is known. 

The masses $m_1$ -- $m_3$ in (\ref{Npsd}) can thus be identified with 
the pole positions found in the 3-pole fits to the electromagnetic 
form factors (in particular, $m_1 = m_\omega$ and $m_2 = m_\phi$). 
Furthermore, the four couplings $g(V,J)$  ($V= \omega, \phi;\, J = 
J^{(I=0)}, \, J^{(s)} \equiv \bar{s} \gamma s$) of the vector meson 
states
\be
\ket{\omega } &=& \cos\epsilon\,\frac{1}{\sqrt{2}} \left(\ket{
\ubar\gamma_\mu u} + \ket{\dbar\gamma_\mu d} \right) - 
\sin\epsilon\, \ket{ \sbar \gamma_\mu s }, \nonumber \\*[7.2pt]
\ket{\phi} &=& \sin\epsilon\,\frac{1}{\sqrt{2}} \left( 
\ket{\ubar\gamma_\mu u} + \ket{\dbar\gamma_\mu d} \right) + 
\cos\epsilon\, \ket{ \sbar \gamma_\mu s } , \label{states}
\ee
(the small angle $\epsilon=0.053$ (Jain {\it et al.}, 1988) 
parametrizes the deviation from ideal mixing) to the neutral 
currents (defined via  $\bra{0} \, J_\mu \,\ket{ V} \, = g(V,J) \,\, 
m^2_V  \, \varepsilon_\mu$)  
are related by the assumption that the quark current of flavor $i$ 
couples to the flavor-$j$ component of the vector meson $V$ with 
universal strength $\kappa$, and only for $i=j$, {\it i.e.}
\be
\bra{0} \, \qbar_i \gamma_\mu  q_i \,\ket{\, (\qbar_j q_j)_V} \, = 
\kappa \, m^2_V \, \delta_{i j} \, \varepsilon_\mu, \label{me}
\ee
which works very well for the electromagnetic couplings. After 
parametrizing the vector-meson nucleon couplings as $g_i (\omega_0, 
N) = g_i \cos \eta_i$, $\, g_i(\phi_0,N) =  g_i \sin \eta_i$ ($i = 
1,2$ denote the $\gamma_\mu$ and $\sigma_{\mu \nu} q^\nu$ 
couplings and $\omega_0, \phi_0$ the ideally mixed states) the four 
couplings $B_{1,2}^{(\omega, \phi)}$ in (\ref{Npsd}) can be obtained 
from the corresponding (fitted) isoscalar couplings $A_{1,2}^{(\omega, 
\phi)}$, which determine phenomenological values for $\eta_i$ and 
$\kappa g_i$:
\be
A_i^{(\omega)} =& \frac{\kappa}{\sqrt{6}} \sin (\theta_0 + 
\epsilon) g_i \cos (\eta_i + \epsilon), \qquad
B_i^{(\omega)} &= - \kappa \sin \epsilon g_i \cos (\eta_i + 
\epsilon),  \label{omegcoupl} \\
A_i^{(\phi)} =& - \frac{\kappa}{\sqrt{6}} \cos (\theta_0 + 
\epsilon) g_i \sin (\eta_i+ \epsilon), \qquad
B_i^{(\phi)} \,\, &= \kappa \cos \epsilon g_i \sin (\eta_i+ \epsilon).
\label{phicoupl}
\ee
($\theta_0$ is the ``magic angle'' with $\sin^2 \theta_0 = 1/3$.) Since 
the flavor content of the strength associated with $m_3$ is unknown, 
the above strategy cannot be applied to the couplings $B_i^{(3)}$. 
They are fixed instead by imposing weak constraints on the 
asymptotic behavior, 
\be
\lim_{q^2 \rightarrow - \infty}  \,\, q^{2(i-1)} F_i^s(q^2) \rightarrow 
0 \qquad \Rightarrow \qquad \sum_v B_1^{(v)} = 0,  \quad \sum_v B_2^{(v)} 
m_v^2 = 0, \label{as1} 
\ee
which also normalize $F_1$. Jaffe's analysis included the minimal 
number of 3 poles in (\ref{Npsd}) and took the couplings $A_i^{(v)}$ 
from the almost twenty year old fits 8.1, 8.2 and 7.1 of 
H{\"o}hler {\it et al.} (1976) (with $m_3 = \{1.4, 1.8, 1.67 \} 
{\rm GeV}$, respectively). The resulting moments, averaged over the 
fits, were $r_s^2 = (0.16 \pm 0.06) \, {\rm fm}^2$, $(r_s^2)_{Sachs} = 
(0.14 \pm 0.07) \, {\rm fm}^2$ and $\mu_s = - (0.31 \pm 0.09)$. 
The indicated error estimates originate solely from the spread between 
the fits and are thus at best a rough lower bound on the complete error.

A new 3-pole fit to the current world data set of the electromagnetic 
form factors (Mergell {\it et al.}, 1995) has prompted our update 
(Forkel, 1995) of the strange form factor analysis. Besides being 
based on a considerably expanded data set, the fits of Mergell {\it 
et al.} are designed to reproduce the logarithmic QCD corrections to 
the form factor asymptotics, which partially originate from continuum 
contributions. Also, they find the third pole mass at the value of 
another well established resonance in the isoscalar channel, $m_3 = 
1.6 \,{\rm GeV}$. The strange form factor analysis benefits from 
these additional features, since they increase the reliability of the 
extracted mass and coupling parameters. The updated values of the 
strangeness radius and magnetic moment are (Forkel, 1995)
\be
r_s^2 = 0.22 \,{\rm fm}^2, \qquad \,\, (r_s^2)_{Sachs} = 0.20 \, 
{\rm fm}^2, \qquad \,\, \mu_s = -0.26. 
\label{mom}
\ee
While the square radius becomes 40 \% larger than that found by 
Jaffe, $| \mu_s |$ is reduced by about 20 \%. The bulk of the 
changes in $r_s^2$ and $\mu_s$ can be traced to differences in the 
$\phi$-nucleon couplings of the used fits. Note that the estimates 
(\ref{mom}) are surprisingly large, of the order of the neutron 
charge radius $r^2_n = -0.11 \, {\rm fm}^2$ and the isoscalar 
magnetic moment of the nucleon, $\mu^{(I=0)} = 0.44$, respectively. 

The momentum dependence of the 3-pole form factors at spacelike $Q^2 
\equiv - q^2$ is shown in Figs. 1a and 1b. It reflects the sizes of 
the pole couplings: since the $| B_i^{(\omega)} |$ are about an order 
of magnitude smaller than the $| B_i^{(\phi)} |$, the leading $1/q^2$ 
dependence of the $\phi$ pole cannot be cancelled (as required by 
(\ref{as1})) by the $\omega$ pole alone. Thus the coupling of 
the third pole must be of similar magnitude as the $\phi$ coupling, 
but of opposite sign. One therefore expects a dipole form of $F^s_2$ 
with a mass parameter between $m_2$ and $m_3$, and an almost perfect 
fit for all space-like momenta is indeed obtained by the simple 
parametrization
\be
F_1^{(s)} (q^2) = \frac16 \frac{r_s^2 q^2}{( 1-\frac{q^2}{M_1^2} )^2},  
\qquad F_2^{(s)} (q^2) = \frac{\mu_s}{( 1-\frac{q^2}{M_2^2} )^2},
\label{dip23p}
\ee
with $M_1 = 1.31\, {\rm GeV} \simeq M_2 = 1.26 \,{\rm GeV}$ (for 
Mergell {\it et al.}s parameters, i.e. with $r_s^2= 5.680 \, {\rm 
GeV}^{-2}$ and $\mu_s = -0.257$), which explicitly realizes the 
asymptotic behavior (\ref{as1}).

The 3-pole ansatz cannot, however, reproduce the asymptotic power 
behavior of the form factors established via QCD dimensional counting 
rules\footnote{I am indebted to Stan Brodsky for helpful correspondence 
on this issue.} (Brodsky and Farrar, 1975; Lepage and Brodsky, 1980). 
Ultimately, at very large, spacelike $q^2$, the form factors are 
dominated by extrinsic contributions, which originate from the 
renormalization of the strange current, are thus suppressed by higher  
powers of $\alpha$, and decay as
\be
F_1^{(s)} (q^2) \rightarrow \left(- q^2 \right)^{-2}, \qquad
F_2^{(s)} (q^2) \rightarrow \left(- q^2 \right)^{-3}.
\label{asymex}
\ee 
However, enforcing this behavior might not necessarily be the best 
choice for an optimal description of the form factors at {\it small 
and intermediate} momentum transfers in the pole approximation. The 
reason is that the other, intrinsic contributions, which originate 
from $s \bar{s}$ admixtures to the nucleon wave function, are,
although asymptotically subleading,
\be
F_1^{(s)} (q^2) \rightarrow \left(- q^2 \right)^{-4}, \qquad 
F_2^{(s)} (q^2) \rightarrow \left(- q^2 \right)^{-5},
\label{asymin}
\ee  
not $\alpha$-suppressed. There might thus exist an intermediate range 
of momentum transfers where the form factors show the intrinsic decay 
behavior. Up to these momenta, the form factors would then be better 
described by enforcing the intrinsic behavior. Furthermore, the pole 
approximation is more reliable at smaller momenta, where also the 
deviation from the extrinsic behavior in the asymtotic tail would be 
of little effect. In the following we will briefly discuss the two 
minimal (4- and 6-pole) ans{\"a}tze which can describe the extrinsic 
or intrinsic asymptotics (for more details see (Forkel, 1995)). In the 
framework of eq. (\ref{Npsd}) the extrinsic power behavior (\ref{asymex}) 
requires minimally 4 poles,
\be
F_i^{(s)}(q^2) = \sum_{v=1}^4 \, B_i^{(v)} \, \frac{m^2_v}{m^2_v - 
q^2}, \qquad i \in \{1,2\},
\label{f4p}
\ee
together with the normalization and asymptotic constraints 
\be 
\left(\begin{array}{cc} m_3^2 & m_4^2 \\ m_3^4 & m_4^4 \end{array} 
\right)
\left(\begin{array}{cc}  m_3^{-2} & 0  \\ 0 & m_4^{-2} \end{array} 
\right) 
\left(\begin{array}{c} B_1^{(3)} \\ B_1^{(4)} \end{array} \right) = 
&-& \left(\begin{array}{c} C_1^{(3)} \\ C_1^{(4)} \end{array} 
\right), \label{scr1matrix} \\
\left(\begin{array}{cc} m_3^2 & m_4^2 \\ m_3^4 & m_4^4 \end{array} 
\right) 
\left(\begin{array}{c} B_2^{(3)} \\ B_2^{(4)} \end{array} \right) = 
&-& \left(\begin{array}{c} C_2^{(3)} \\ C_2^{(4)} \end{array} \right), 
\label{scr2matrix}
\ee
($C_1^{(i)} \equiv \sum_{j=1}^2 B_1^{(j)} m_j^{2(i-3)}$, $C_2^{(i)} 
\equiv \sum_{j=1}^2 B_2^{(j)} m_j^{2(i-2)}$), which have a unique 
solution for  the couplings $B_i^{(3,4)}$ as a function of the masses 
$m_{3,4}$. These solutions leave the value of the fourth mass, $m_4$, 
free. Maintaining the third pole at $m_3 = 1600 \,{\rm MeV}$ and 
requiring $m_4$ to be larger than $m_3$ by at least a typical width 
of $ \sim 300 \, {\rm MeV}$ (so that it can be resolved in zero-width 
approximation), i.e. $m_4 \geq 1.9  \,{\rm GeV}$, the 
results for the strangeness radius and magnetic moment interpolate 
smoothly and monotonically in the range 
\be 
0.15 \, {\rm fm}^2 \le &r_s^2& \le 0.22 \, {\rm fm}^2, \nonumber \\  
0.14 \, {\rm fm}^2 \le &\,\, (r_s^2)_{Sachs}\,\,& \le 0.20 \, 
{\rm fm}^2,  \nonumber \\
-0.18 \ge &\mu_s& \ge -0.26.
\label{4pbounds}
\ee
For $m_4 \rightarrow \infty$ the fourth pole does affect the momentum 
dependence only at $Q^2 >> m_4^2$, and for smaller $Q^2$ the 4-pole 
ansatz becomes identical to the 3-pole ansatz, which provides the 
upper bounds on $r_s^2$ and $| \mu_s |$ in (\ref{4pbounds}). A fourth 
pole in the $2 \,{\rm GeV}$ region, however, reduces the 3-pole moments  
by about a third. For all admissable values of $m_4$, the couplings 
of the third pole are necessarily large. In the 4-pole ansatz with 
$m_4 \sim 2 \,{\rm GeV}$, also the coupling of the fourth pole is 
of comparable size and we expect quadrupole form factors. 
Indeed, the conservative choice $m_4=1.9 \, {\rm GeV}$ is well 
fitted by $F_1^{(s)}(q^2) = (r_s^2 q^2 /6)( 1- q^2/M_1^2 )^{-3}, 
\, F_2^{(s)}(q^2) = \mu_s (1 - q^2/M_2^2 )^{-3} $ with $r_s^2 = 
0.1482 \,{\rm fm}^2, \mu_s = -0.1789,  M_1 = 1.47 \,{\rm GeV}$ and 
$M_2 = 1.43 \,{\rm GeV}$, which provide a lower bound on the 
(absolute) magnitude of the form factors with extrinsic asymptotics. 

We now turn to the intrinsic asymptotics (\ref{asymin}), which 
requires two more superconvergence relations for each form factor and 
minimally 6 poles, since the 5-pole ansatz would be overconstrained 
(Forkel, 1995). The relevant expressions are direct 
generalizations of (\ref{f4p}), (\ref{scr1matrix}) and (\ref{scr2matrix}) 
from $N=4$ to $N=6$. As for the 4-pole ansatz, the couplings can be 
uniquely expressed in terms of the masses, leaving two more pole 
positions, $m_5$ and $m_6$, undetermined. Again we estimate the range 
of possible 6-pole form factors by requiring spacings of minimally $300 
\, {\rm MeV}$ between higher-lying poles. The most conservative estimate 
corresponds then to the mass values ($\{m_4, m_5, m_6 \} = \{1.9, 
2.2, 2.5 \} \,{\rm GeV}$) and is again well fitted by the 
simplest one-mass-parameter formulae which match their asymptotic 
behavior: $F_1^{(s)}(q^2) = (r_s^2 q^2/6)( 1-q^2/M_1^2)^{-5} , 
\,F_2^{(s)} (q^2) = \mu_s ( 1- q^2/M_2^2)^{-5}$, with $r_s^2 = 
0.08879 \,{\rm fm}^2, \mu_s = -0.1136,  M_1 = 1.72 \,{\rm GeV}$ (or 
somewhat better for $Q^2 \le 10 \,{\rm GeV}$ by $M_1 = 1.79 \,{\rm 
GeV}$) and $M_2 = 1.66 \,{\rm GeV}$. Increasing $m_6, m_5$ and $m_4$ 
up to infinity one arrives again at the 3-pole form factors as an 
upper bound. The range of values for the leading moments is now 
given by 
\be 
0.089 \, {\rm fm}^2 \le &r_s^2& \le 0.22 \, {\rm fm}^2, \nonumber \\  
0.081 \, {\rm fm}^2 \le & \,\,(r_s^2)_{Sachs} \,\,& \le 0.20 \, 
{\rm fm}^2, \nonumber \\  -0.086 \ge &\mu_s& \ge -0.26.
\label{6pbounds}
\ee
The intrinsic asymptotics can thus reduce the size of the strangeness 
radius and magnetic moment by about a factor of 3 relative to that of 
the 3-pole estimate. Comparing the different asymptotics we 
arrive at some general conclusions: (i) the $r_s^2$- and 
$|\mu_s|$--values of the minimal 3-pole ansatz are upper bounds and 
very likely overestimated, possibly by up to a factor of three, (ii) 
their values contain (at least in the realm of the pole approximation) 
significant information on the asymptotic behavior and its onset, and 
(iii) the pole approximation leads quite generally to {\it positive} 
strangeness radii and negative magnetic moments. The last point can be 
readily understood from the generic $N$--pole expressions 
\be
r_s^2 = \sum_v^N \frac{ B_1^{(v)} }{m_v^2}, \qquad \qquad \mu_s =
 \sum_v^N  B_2^{(v)}, \label{momexpl}
\ee
since the large $\phi$ couplings are positive in $F_1^{(s)}$ and 
negative in $F_2^{(s)}$ and since both the alternating signs of the 
couplings (due to the superconvergence relations) and the $m_v^{-2}$ 
factor in $r_s^2$ suppress higher-pole contributions. The sign of the 
strangeness radius might, however, be changed by ({\it e.g.} $K 
\bar{K}$) continuum contributions, as discussed in the following 
sections. 

\vskip 2\baselineskip
\hskip 2.5cm GENERALIZED VECTOR MESON DOMINANCE

The pole approximation of the above dispersive analysis includes 
effects of the $\bar{K} K$ continuum (and those from the other cuts) 
at best implicitly. In the present section we present a model for the 
strange form factors which is based on a more general version of VMD 
and contains explicit contributions from the kaon cloud of the nucleon. 
This approach relies exclusively on the lightest, narrow isoscalar 
vector mesons, $\omega$ and $\phi$, and is discussed in detail by Forkel 
{\it et al.} (1994). The VMD hypothesis is formulated in terms of current 
field identities (CFIs) (Kroll, Lee and Zumino, 1967), which imply the 
proportionality of the electromagnetic current to the field operators 
of the light, neutral vector mesons with the same quantum numbers. In 
the isocalar electromagnetic channel the CFI reads
\be
J_\mu^{(I=0)} = A_\omega \, m_\omega^2 \, \omega_\mu + A_\phi 
\, m_\phi^2 \, \phi_\mu, \label{cfi1}
\ee
with the couplings  $A_\omega$, $A_\phi$ yet to be fixed. Generalizing 
VMD to the strangeness current, we write an analogous CFI
\be
J_\mu^{(s)} = \bar{s} \gamma_\mu s = B_\omega \, m_\omega^2 \, 
\omega_\mu + B_\phi \, m_\phi^2 \, \phi_\mu. \label{cfi2}
\ee
After combining eqs. (\ref{cfi1}) and (\ref{cfi2}) into a vector 
equation, the couplings form the elements of a matrix $\hat{C}$. 
Sandwiching the CFIs between the physical vector meson states and 
the vacuum, and using eq. (\ref{me}), we obtain an explicit form 
for $\hat{C}$,
\be
\hat{C}_{I=0,s}(\epsilon) = \left(\begin{array}{cc}
A_\omega & A_\phi \\ B_\omega & B_\phi \end{array} \right) = 
\kappa \, \left(\begin{array}{cc}
{1\over\sqrt{6}}\sin(\theta_0 + \epsilon) & 
{-1\over\sqrt{6}} \cos(\theta_0 + \epsilon)\\
-\sin\epsilon & \cos\epsilon
\end{array} \right) .
\label{C}
\ee
The CFIs lead to a general expression for the form factors. To 
derive it, we first note that eqs. (\ref{cfi1}) and (\ref{cfi2}), 
together with the requirement of strangeness and hypercharge
conservation, imply $\partial^\mu V_\mu = 0$ ($V_\mu$ stands 
for either $\omega_\mu$ or $\phi_\mu$), which simplifies the 
field equations to 
\be
( \Box + m^2_V ) \, V_\mu = J_\mu^{(V)} \label{feq}
\ee
and therefore also implies that the vector meson source currents 
are conserved ($\partial^\mu J_\mu^{(V)} = 0$). We now take 
nucleon matrix elements of the field equations (\ref{feq}) and 
use the CFI's to write 
\be
\left(\begin{array}{c}\bra{N(p')} \, J_\mu^{(I=0)} \,\ket{ N(p)}\\
\bra{N(p')} \, J_\mu^s \,\ket{ N(p)}\end{array} \right)\ =  
\hat{C}_{I=0,s}(\epsilon)  \left(\begin{array}{cc}
\frac{m^2_\omega}{m_\omega^2-q^2} & 0\\
0 & \frac{m_\phi^2}{m_\phi^2-q^2}
\end{array} \right)\ 
\left(\begin{array}{c}\bra{ N(p') }\, J_\mu^{(\omega)} \,\ket{ N(p)}   
\\\bra{N(p')} \, J_\mu^{(\phi)} \,\ket{N(p)}\end{array} \right).
\label{feq2}
\ee
Is is convenient to reexpress the vector meson source currents as 
linear combinations of currents with the same $SU(3)$ transformation 
behavior as $J^{(I=0)}$ and $J^{(s)}$, which we will denote as {\it 
intrinsic} ($J_{in}$). After furthermore separating the nucleon matrix 
elements into form factors, according to eq. (\ref{sff}) and its analog 
for $J_\mu^{(I=0)}$, we obtain our general VMD expression for the form 
factors:
\be
\left(\begin{array}{c}
F^{I=0}(q^2)\\
F^{(s)}(q^2)
\end{array} \right) = \hat{C}_{I=0,s}(\epsilon)
\left(\begin{array}{cc}
\frac{m^2_\omega}{m_\omega^2-q^2} & 0\\
0 & \frac{m_\phi^2}{m_\phi^2-q^2}
\end{array} \right)\hat{C}_{I=0,s}^{-1}(\epsilon)
\left(\begin{array}{c}
F_{in}^{I=0}(q^2)\\
F_{in}^{(s)}(q^2)
\end{array}\right)\ .
\label{form1}
\ee
According to their definition, the intrinsic form factors describe
the extended source current distribution of the nucleon to which the 
vector mesons couple. Since both $J^{(I=0)}$, $J^{(s)}$ and their 
intrinsic counterparts  $J_{in}^{(I=0)}$, $J_{in}^{(s)}$
are conserved, the full and the intrinsic form factors in eq. 
(\ref{form1}) have the same normalization at $q^2 = 0$. Combining 
eqs. (\ref{C}) and (\ref{form1}), we finally obtain
\be
\left(\begin{array}{c}
F^{I=0}(q^2)\\
F^{(s)}(q^2)
\end{array} \right) & = &
\left(\begin{array}{cc}
\frac{m^2_\omega}{m_\omega^2-q^2}\frac{\sin(\theta_0+\epsilon)
\cos\epsilon}{\sin\theta_0}-\frac{m^2_\phi}{m_\phi^2-q^2}
\frac{\cos(\theta_0+\epsilon)\sin\epsilon}{\sin\theta_0} & 
\frac{\cos(\theta_0+\epsilon)\sin(\theta_0+\epsilon)}{\sqrt{6}
\sin\theta_0}\left( \frac{m^2_\omega}{m_\omega^2-q^2}-\frac{m_\phi^2}
{m_\phi^2-q^2}  \right) \\
{\sqrt{6}\cos\epsilon\sin\epsilon\over\sin\theta_0}
\left( \frac{m_\phi^2}
{m_\phi^2-q^2} - \frac{m^2_\omega}{m_\omega^2-q^2}\right)  &  
\frac{m_\phi^2}{m_\phi^2-q^2} 
{\cos\epsilon\sin(\theta_0+\epsilon)\over\sin\theta_0} - 
\frac{m_\omega^2}{m_\omega^2-q^2}{\sin\epsilon\cos(\theta_0
+\epsilon)\over\sin\theta_0}  
\end{array} \right) 
\nonumber\\*[7.2pt]
& &\times\left(\begin{array}{c}
F_{in}^{I=0}(q^2)\\
F_{in}^{(s)} (q^2)
\end{array}\right).
\label{form}
\ee
Note that eq. (\ref{form}) is independent of the overall 
vector-meson-current and vector-meson-nucleon coupling constants, 
which cancel each other due to charge normalization. 

Up to now our discussion has been rather general, and different 
choices for the intrinsic form factors can be implemented in the given 
framework, as long as they do not lead to double counting with the VMD 
sector. Here, we adopt the kaon loop model of Musolf and Burkhardt  
(1993) for the intrinsic strangeness form factor (but use the physical 
value for the $\Lambda$ mass instead of their flavor-symmetric value). 
This model describes the current-nucleon vertex corrections due to 
$K$--$\Lambda$ loop graphs. Although the latter are U.V. finite, 
the loop momenta are cut off by 
meson-nucleon vertex form factors $H(k^2) = (m_K^2 - \Lambda^2)/(k^2 
- \Lambda^2)$ from the Bonn potential (Holzenkamp {\it et al.}, 1989),
since the effective hadronic description of the underlying physics 
breaks down at large momenta. The Bonn values for the cutoff $\Lambda$ 
in the $N\Lambda K$ vertex were extracted from fits to baryon-baryon 
scattering data and lie in the range of 1.2 -- 1.4 $\GeV$. One  
finds three amplitudes, $\Gamma_\mu^{(B,M,W)}$, which contribute to 
the intrinsic form factors. They are associated with processes where 
the current couples either to the baryon line (B), the meson line (M) 
or the meson-baryon vertex (V) in the loop, and the intrinsic strange 
form factors are obtained from the nucleon matrix element of their 
sum,
\be
\overline{N}(p^\prime) \left[ \Gamma^B_\mu(p^\prime,p) + 
\Gamma^M_\mu(p^\prime,p) + \Gamma^V_\mu(p^\prime,p) \right] N(p) 
= \overline{N}(p^\prime)\left(\gamma_\mu F_{1,in}^{(s)}(q^2) +
 i\frac{\sigma_{\mu\nu}q^\nu} {2M_N} F_{2,in}^{(s)} (q^2)
\right)N(p) .
\ee
Explicit expressions for the $\Gamma^{(B,V,M)}$ are given in 
(Forkel {\it et al.}, 1994), where it is also shown that their sum 
satisfies the Ward-Takahashi identity. The values 
of the coupling and masses are fixed at $M_N=939\MeV$, $M_\Lambda=1116 
\MeV$, $m_K=496\MeV$ and $g_{N\Lambda K}/ \sqrt{4\pi}=-3.944$ 
(Holzenkamp {\it et al.}, 1989). Finally, we extract the {\it 
intrinsic isoscalar} form factors from the fit to the measured 
isoscalar form factors by inverting the VMD matrix in 
eq.~(\ref{form}). The full strangeness form factors are then 
determined by the second row of eq.~(\ref{form}). (The contribution 
from the intrinsic strangeness part to the isoscalar form factor is 
very small and plays almost no role in the determination of 
$F_{in}^{I=0}(q^2)$.)

In the generalized VMD approach the strangeness magnetic moment 
originates solely from the intrinsic kaon loop contribution 
({\it cf.} eq. (\ref{form1})), while both the Dirac and the Sachs 
strangeness radii get an additive contribution from the vector mesons. 
One finds 
\be
r_{s}^2 = - (0.0243 - 0.0245) \, {\rm fm}^2, \quad 
(r_{s}^2)_{Sachs} = - (0.040 - 0.045) \,  {\rm fm}^2, \quad
\mu_s = - (0.24 - 0.32). 
\ee
The signs of the strangeness radii are the same as those of the 
intrinsic contribution. 

The momentum dependence of the resulting Dirac and Pauli form factors 
is shown in Figs. 2a and 2b. For comparison, we also show the 4-pole 
and 6-pole form factors of Figs. 1a and 1b. The two approaches differ 
in sign and magnitude of the Dirac form factor (and thus of the 
strangeness radius), but yield more similar Pauli form factors. 
(The Pauli form factor from generalized VMD is almost identical to 
that of the 3-pole ansatz.) The combined 
data from SAMPLE (which measures $F_2^{(s)}$ at $Q^2 = 0.1 \GeV^2$), 
CEBAF (in particular the G0 experiment (Beck {\it et al.}, 1991), 
which will measure  $G_M^s$ in the momentum range $0.1 \GeV^2 <Q^2< 
0.5 \GeV^2$ with a resolution $\delta\mu_s \simeq \pm0.22$ at low 
$Q^2$) and MAMI (Heinen-Konschak {\it et al.}, 1993) should thus be 
sufficient to distinguish between the two VMD approaches. Note, 
finally, that the strangeness radius from generalized VMD is 
proportional to the sine of the mixing angle $\epsilon$ and would 
thus not receive any contribution from ideally mixed vector meson 
states, whereas in the dispersion analysis $r^2_s$ gets bigger as 
the mixing angle $\epsilon$ goes to zero, since the overall strangeness 
of the intrinsic charge distribution vanishes. 

\vskip 2\baselineskip
\hskip 2.5cm SUMMARY AND CONCLUSIONS
\label{sum}

The vector meson dominance mechanism has a largely generic character 
and successfully describes electromagnetic interactions of hadrons. 
It should therefore be a useful starting point for estimates of the 
nucleon's strange vector form factors. We have discussed two such 
estimates: the first, a dispersive treatment, relies 
on phenomenological input and is nucleon-model independent, while the 
second builds on current field identities and consistently includes 
an intrinsic strangeness distribution due to the nucleon's kaon cloud. 

The dispersive analysis, which we restrict to the pole approximation, 
is based on input from the isoscalar electromagnetic form factor data. 
We show that the minimal parametrization of the spectral functions, 
Jaffe's 3-pole ansatz, yields upper bounds for the magnitude of the 
strangeness radius $r_s^2$ and magnetic moment $ \mu_s $. Using new 
fits to the current world data set of the isoscalar form factors, we 
also update the results of the 3-pole analysis and find that 
$r_s^2$ increases by 40 \% while $| \mu_s|$ decreases by 20 \%, 
compared to Jaffe's original values. Due to the unrealistic asymptotics 
of the 3-pole ansatz, however, these upper bounds do probably  
overestimate the moments: the reproduction of the asymptotic behavior 
derived from QCD counting rules requires additional poles and leads to 
significantly reduced values for the moments. Implementing the leading 
QCD asymptotics with a fourth pole term and a conservative estimate for 
the bulk position of higher-lying strength reduces the 3-pole results  
by more than a third, to $r_s^2 = 0.15 \, {\rm fm}^2, \; (r_s^2)_{Sachs} 
=0.14 \, {\rm fm}^2, \; \mu_s = -0.18$, while the stronger decay due 
to intrinsic contributions leads to a further reduction by up to 
50 \%,  $r_s^2 = 0.089 \, {\rm fm}^2, \; (r_s^2)_{Sachs} 
=0.081 \, {\rm fm}^2, \; \mu_s = -0.086$.

Due to the alternating signs (``bump-dip'' structure) of neighboring 
pole couplings, the momentum dependence of the pole-ansatz form factors 
turns out to be well fitted by simple multipole parametrizations. The 
positive sign of $r_s^2$ and the negative sign of $\mu_s$ follow the 
signs of the large $\phi$ couplings and are thus generic in the pole 
framework. The fact, furthermore, that the lightest relevant (i.e. 
$K$--$\Lambda$) intermediate states in the nucleon wave function 
produce the opposite, 
negative sign for $r_s^2$ may indicate that the $K$--$\bar{K}$ 
continuum could change the sign of the strangeness radius in the 
dispersive analysis. 

This sign change indeed takes place in a generalization of the VMD 
framework which consistently includes kaonic contributions on the 
basis of current field identities. The latter permit to account 
explicitly for an intrinsic strangeness distribution of the nucleon, 
which we generate by a $K$--$\Lambda$ loop amplitude. Restricting 
the VMD sector to the sharp $\omega$ and $\phi$ resonances, we find a 
negative and by half an order of magnitude smaller strangeness radius 
($r_{s}^2 = - 0.024 \, {\rm fm}^2, \, (r_{s}^2)_{Sachs} \approx - 
0.043 \,  {\rm fm}^2$), while the strange magnetic moment predictions 
are of the same sign and a similar (loop-cutoff dependent) magnitude 
than those of the dispersion analysis, $\mu_s = - (0.24 - 0.32)$.

Both the dispersive and the CFI-based analysis allow many refinements. 
Perhaps the most important ones are the inclusion of $K \bar{K}$ 
continuum contributions in the former and a more complete model for 
the intrinsic form factors (Forkel, Musolf and Nielsen, 1996) in the 
latter. Investigations in both directions are in progress. 

I would like to thank Professor Amand Faessler and the 
co-organizers of the school for creating such a stimulating, 
well-rounded and productive meeting. I would also like to thank 
Tom Cohen, Xuemin Jin and Marina Nielsen for the enjoyable 
collaboration on the generalized VMD approach and many colleagues, 
including Mary Alberg, Betsy Beise, Stan Brodsky, Gerhard Ecker, 
Klaus Goeke, Gottfried Holzwarth, Fritz Klein, Mike Musolf, Antoni 
Szczurek and Wolfram Weise for discussions. This work was supported 
by the U.S. Department of Energy under Grant No.\ DE--FG02--93ER--40762 
and by the European Community through the HCM programme.


\vskip 2\baselineskip
\hskip 2.5cm REFERENCES

Ahrens, L.A. {\it et al.} (1987). {\it Phys.\ Rev.} {\bf D35}, 785. \\
Alberg, M. (1995). Contribution to this volume. \\
Ashman, J. {\it et al.} (1988). {\it Phys.\ Lett.} {\bf B206}, 364. \\
Ashman J. {\it et al.} (1989). {\it Nucl. Phys.} {\bf B328}, 1. \\
Beck, D.H. {\it et al.} (1991). CEBAF proposal \#PR-91-017.\\
Brodsky, S.J. and G.R. Farrar (1975). {\it Phys. Rev.} {\bf D11}, 
1309. \\
Cheng, T.P. and R. Dashen (1971). {\it Phys. Rev. Lett.} {\bf 26}, 
594.\\
Cheng, T.P. (1976). {\it Phys. Rev.} {\bf D13}, 2161.\\
Cohen, T.D. , H. Forkel and M. Nielsen (1993). {\it Phys. Lett.} 
{\bf B316}, 1. \\
Donoghue, J.F. and C.R. Nappi (1986). {\it Phys.\ Lett.} {\bf B168}, 
105.\\
Ellis, J. and R.L. Jaffe (1974). {\it Phys. Rev.} {\bf D9}, 1444.\\
Forkel, H. {\it et al.} (1994). {\it Phys. Rev.} {\bf C50}, 3108.\\
Forkel, H. (1995). Preprint ECT*/Sept/014-95. \\
Forkel, H., M.J. Musolf and M. Nielsen (1996). In preparation. \\
Garvey, G.T., W.C. Louis and D.H. White (1993). {\it Phys. Rev.} 
{\bf C48}, 761.\\
Gasser, J., H. Leutwyler and M.E. Sainio (1991). {\it Phys.\ Lett.} 
{\bf B253}, 252.\\
Genz, H. and G. H{\"o}hler (1976). {\it Phys. Lett.} {\bf B61}, 389. 
\\
Heinen-Konschak, E. {\it et al.} (1993). MAMI proposal A4/1-93. \\
H{\"o}hler G. {\it et al.} (1976). {\it Nucl.\ Phys.} {\bf B114}, 505. 
\\
Holzenkamp, B. {\it et al.} (1989). {\it Nucl.\ Phys.} {\bf A500}, 485. 
\\
Ioffe, B.L. and M. Karliner (1990). {\it Phys. Lett.} {\bf B247}, 387. 
\\
Jaffe, R.L. (1989). {\it Phys.\ Lett.} {\bf B229}, 275. \\
Jain, P. {\it et al.} (1988). {\it Phys.\ Rev.} {\bf D37}, 3252. \\
Kaplan, D.B. and A. Manohar (1988). {\it Nucl. Phys.} {\bf B310}, 527. 
\\
Kluge, W. (1995). Contribution to this volume. \\
Kroll, N.M., T.D. Lee and B. Zumino (1967). {\it Phys. Rev.} {\bf 157}, 
1376.\\
Lepage, G.P. and S.J. Brodsky (1980). {\it Phys. Rev.} {\bf D22}, 2157. 
\\
Liu, K.F. and S.J. Dong (1994). {\it Phys. Lett.} {\bf B328}, 130.\\
McKeown, R.D. and D.H.Beck (1989). MIT/Bates proposal \# 89-06.\\
Mergell, P., U.-G. Mei{\ss}ner and D. Drechsel (1995). Mainz and 
Bonn University preprint TK 9515, \hglue 0.5 cm MKPH-T-95-07, 
hep-ph/9506375. \\
Musolf, M.J. {\it et al.} (1994). {\it Phys. Rep.} {\bf 239}, 1.\\
Musolf, M.J. and M. Burkardt (1994). {\it Z. Phys.} {\bf C61}, 433.\\

\newpage

\begin{figure}
\caption{The strange a) Dirac and b) Pauli vector form factors from 
the 3-pole (continuous line), 4-pole (dashed line) and 5-pole 
(dotted line) ans{\"a}tze.}
\label{fig-1}
\end{figure}

\begin{figure}
\caption{The a) Dirac and b) Pauli form factors from generalized 
VMD (full line) in comparison with those from the 4-pole (dotted 
line) and 6-pole (dashed line) ansatz.}
\label{fig-2}
\end{figure}

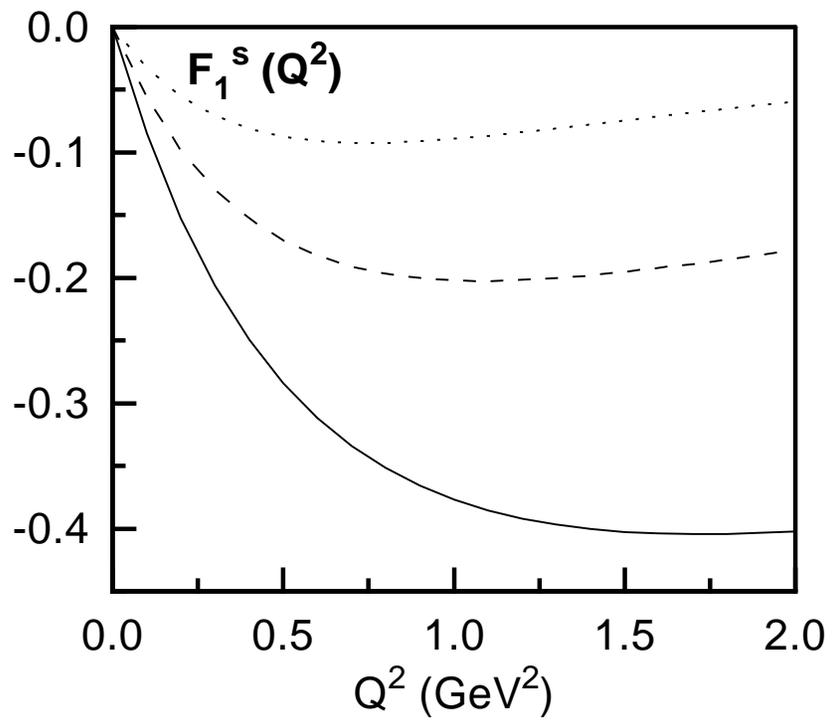
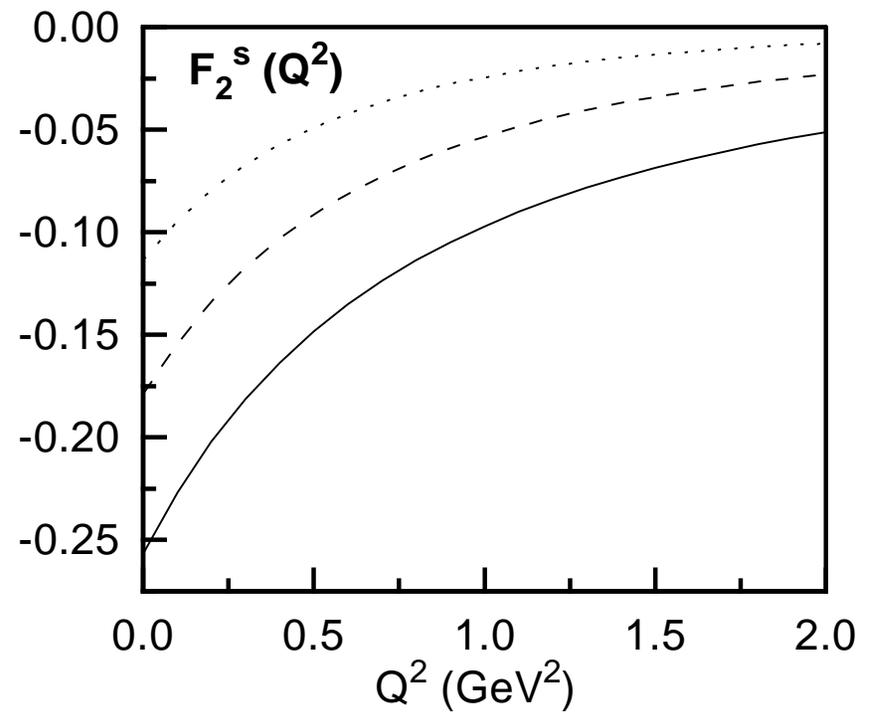
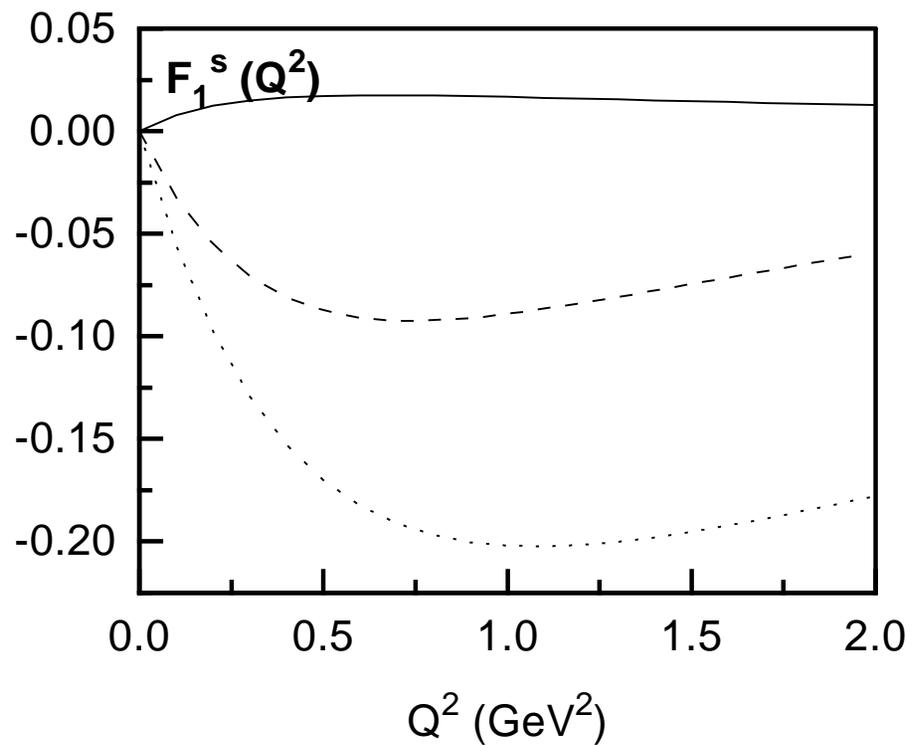
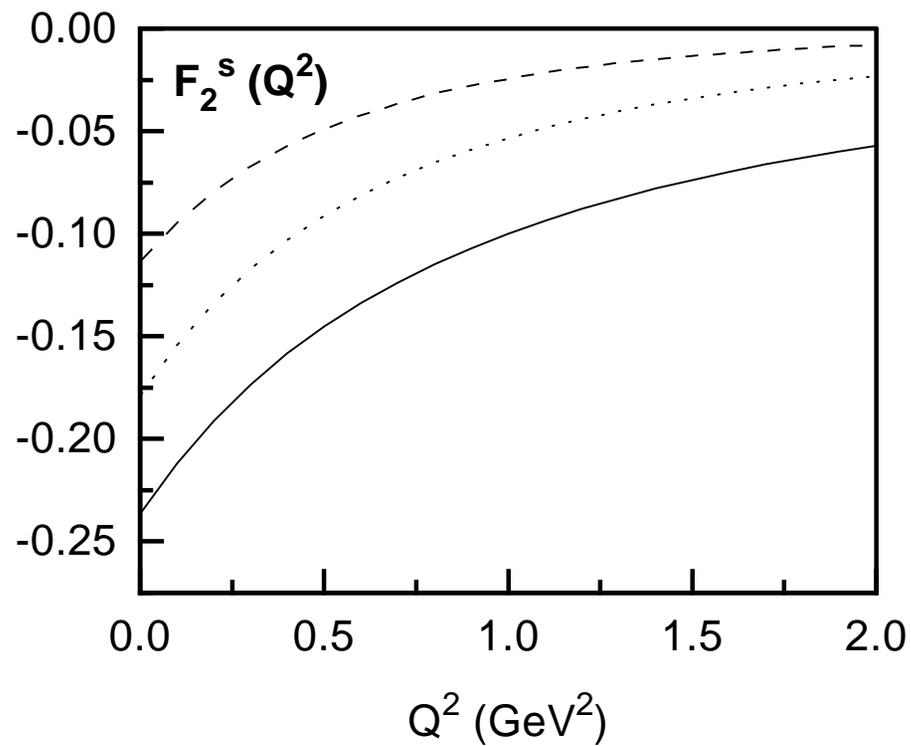
\end{document}